\documentclass[twocolumn]{revtex4-1}
\usepackage[latin9]{inputenc}
\usepackage{amsmath}
\usepackage{amssymb}
\usepackage{graphicx}
\begin{document}
\global\long\def\ket#1{\left|#1\right\rangle }

\global\long\def\bra#1{\left\langle #1\right|}

\global\long\def\braket#1#2{\left\langle #1\left|#2\right.\right\rangle }

\global\long\def\ketbra#1#2{\left|#1\right\rangle \left\langle #2\right|}

\global\long\def\braOket#1#2#3{\left\langle #1\left|#2\right|#3\right\rangle }

\global\long\def\mc#1{\mathcal{#1}}

\global\long\def\nrm#1{\left\Vert #1\right\Vert }

\global\long\def\unit{1\!\!1}

\title{Speed limits in Liouville space for open quantum systems}

\author{Raam Uzdin}
\email{raam@mail.huji.ac.il}

\author{Ronnie Kosloff}

\affiliation{Fritz Haber Research Center for Molecular Dynamics, The Hebrew University
of Jerusalem, Jerusalem 9190401, Israel}
\begin{abstract}
One of the defining properties of an open quantum system is the variation
of its purity in time. We derive speed limits on the rate of purity
change for systems coupled to a Markovian environment. Our speed limits
are based on Liouville space where density matrices are represented
as vectors. This approach leads to speed limits that are always tighter
compared to their parallel speed limits in Hilbert space. These bounds
depend solely on the generators of the nonunitary dynamics and are
independent of the particular states of the systems. Thus, they are
perfectly suited to investigate dephasing, thermalization, and decorrelation
processes of arbitrary states. We show that our speed limits can be
attained and are therefore tight. As an application of our results
we study correlation loss, and the speed of classical and quantum
correlation erasure in multi-particle system. 
\end{abstract}
\maketitle
Determining the maximal rate of evolution of an open system is of
crucial importance in quantum physics. Any quantum system unavoidably
is coupled to external degrees of freedom (the environment) that lead
to loss of phase coherence and/or to thermalization \cite{bre07}.
In most applications, the main challenge is to minimize and slow down
the effect of the environment. In quantum computing \cite{nie00}
and coherent control \cite{sha03}, it is vital to achieve low dephasing
rates in order to obtain the target transformation. On the other hand,
in cooling and pure state preparation it is desirable to speed up
the influence of the surroundings. In addition, in quantum thermodynamics,
a bound on the rate of thermalization of a heat engine with its reservoirs
will limit its cycle time and therefore impose a restriction on the
maximal power output \cite{kos14}. An interaction with the environment
is associated with changes of the system's purity $\mathcal{P}=\text{tr}\rho^{2}$
\cite{bre07}, where $\rho$ is the density operator of the system.
Despite the ability of the purity to quantify the impact of the environment
on the system, to the best of our knowledge no general result on the
maximal rate of purity variation has been obtained so far within the
basic Markovian formalism of open quantum systems.

Bounds on the rate of quantum evolution are useful to assess if a
process can be completed in a given time, without having to explicitly
solve the equations of motion \cite{fle73,bha83,ana90,vai91,uff93,bro03}.
Quantum speed limits, have been studied in \cite{man45,mar98,gio03,pfe93,pfe95,def13,tad13,cam13,def13a,Reznik,RusselNav,Lidar2015,Addeso2016,NonMarkSL}.
Their evaluation is of importance in essentially all areas of quantum
physics where the determination of the minimal time of a process is
of interest \cite{bek81,llo00,gio11,hue12,def10}. When speaking about
speed limits of some quantity $G$ (e.g. purity, entropy, or angle
between states) it is important to distinguish between a bound on
the instantaneous rate of change $\left|\frac{dG}{dt}\right|$ and
a bound on the cumulative change $\left|G(t_{f})-G(t_{i})\right|$
carried out over a time interval $[t_{i},t_{f}]$. The two are related
via 
\begin{equation}
\left|G(t_{f})-G(t_{i})\right|=\left|\int_{t_{i}}^{t_{f}}\frac{dG}{dt}dt\right|\le\int_{t_{i}}^{t_{f}}\left|\frac{dG}{dt}\right|dt.\label{eq: G cumu}
\end{equation}
There is, however, an important difference between instantaneous and
cumulative speed limits. Instantaneous speed limits often use the
state of the system $\rho$ to evaluate $\left|\frac{dG}{dt}\right|$.
In quantum metrology where the optimal state for phase estimation
is needed \cite{tad13,cam13}, the state dependence is very useful.
Other well known speed limits such as the Mandelstam-Tamm bound \cite{man45},
and the Margolus-Levitin bound \cite{mar98} are state dependent.
For open systems state dependent speed limits see \cite{rod11,hut12}.
However for the purpose of bounding cumulative changes, state dependence
speed limit cannot be used. If the integrand in the right hand side
of (\ref{eq: G cumu}) depends on the state it means that $\rho(t)$
must be known in order to calculate the bound. However if $\rho(t)$
is available, it possible to calculate $G(t_{f})-G(t_{i})$ exactly
and directly. Thus, for the speed limit to be useful, we impose the
restriction that the bound on the rate of change must be state-independent
(see \cite{LidarNormActionAndDistanceBound,uzdin100evoSpeed,Lidar2015}
for state-independent bounds) and must be expressed solely in terms
of the generators of motion (e.g. Hamiltonian and Lindblad operators
\cite{bre07}) .

Our aim in this paper is to provide a cumulative state-independent
speed limits for the purity change. After presenting a purity speed
limit in the standard Hilbert space of density matrices, we use Liouville
space (space of density \char`\"{}vectors\char`\"{}) to obtain a speed
limit which is always tighter. Furthermore the Liouville space speed
limit overcomes some intrinsic limitations of the Hilbert space speed
limit. Next, we considerably improve our results by introducing the
purity deviation. Finally, we employ our formalism to derive speed
limits for multi-particle dephasing channels and for decorrelating
channels. Although our speed limits are state-independent and cumulative,
we show that in both cases the speed limits can be attained and are
therefore \textit{tight}.

\section{Purity speed limit in Hilbert space}

We begin by deriving a simple yet limited, purity speed limit in the
density matrix formulation. We consider a possibly driven $N$-level
quantum system with Hamiltonian $H$. The effect of the environment
on the dynamics of the system in the weak coupling limit is described
by a Markovian master equation of the Lindblad-type for the density
matrix of the system\cite{bre07}, 
\begin{eqnarray}
d_{t}\rho & = & i[H,\rho]+L_{t}(\rho)=i[H,\rho]\nonumber \\
 & + & e^{i\phi(t)}(\sum_{k}A_{k}\rho A_{k}^{\dagger}-\{\frac{1}{2}A_{k}^{\dagger}A_{k},\rho\})\label{eq: Lindblad eq.}
\end{eqnarray}
where the operators $A_{k}\in\mathbb{C}^{N\times N}$ describe the
interaction with the environment. For Markovian systems $\phi(t)=0$,
however (\ref{eq: Lindblad eq.}) also describes some non-Markovian
systems (\cite{nonMark1,nonMark2,nonMark3}). Our results are valid
for general $\phi(t)$. Master equations of the form (\ref{eq: Lindblad eq.})
are the tool of choice to investigate the dynamics of systems weakly
coupled to a reservoir in quantum optics and solid state physics \cite{bre07}.

From (\ref{eq: Lindblad eq.}) it follows that $d_{t}\ln\text{tr}(\rho^{2})=2\text{tr}(\rho L_{t}(\rho))/\text{tr}(\rho^{2})$.
Integrating over time and using the triangle inequality, we have 
\begin{eqnarray}
\left|\ln\frac{\mc P(t_{f})}{\mc P(t_{i})}\right| & \le & \int\frac{2\left|\text{tr}(\rho L_{t}(\rho))\right|}{\text{tr}(\rho^{2})}dt.\label{eq: rho purity triang}
\end{eqnarray}
Next, we exploit the fact that $\mc P(t)=\text{tr}(\rho^{2})=\nrm{\rho}_{2}^{2}$,
where $\nrm{\cdot}_{2}$ denotes the standard Hilbert-Schmidt (HS)
norm. An upper bound to (\ref{eq: rho purity triang}) can be derived
with the help of elementary matrix algebra \cite{bha97,Horn}. Combining
the Cauchy-Schwarz inequality, $|\text{tr}(\rho L_{t}(\rho))|\le\nrm{\rho}_{2}\nrm{L_{t}(\rho)}_{2}$,
the triangle inequality together with the submultiplicativity property
of the norm and the master equation , we find $\nrm{\rho}_{2}\nrm{L_{t}(\rho)}_{2}\le2\sum_{k}\nrm{A_{k}}_{2}^{2}\nrm{\rho}_{2}^{2}$.
Inserting this expression into (\ref{eq: rho purity triang}), we
obtain a \char`\"{}norm integral\char`\"{} inequality of type (\ref{eq: G cumu})
for the logarithm of the purity 
\begin{equation}
\left|-\ln\mc P(t_{f})+\ln\mc P(t_{i})\right|\le4\int_{t_{i}}^{t_{f}}\sum_{k}\nrm{A_{k}}_{2}^{2}dt,\label{eq: H_rho purity bound}
\end{equation}
By using $|\text{tr}(\rho L_{t}(\rho))|\le\nrm{\rho}_{2}\nrm{L_{t}(\rho)}_{2}\le\nrm{L_{t}(\rho)}_{2}$
one can also get a speed limit for $\mc P$ rather than for $-\ln\mc P$
but it will be less tight than (\ref{eq: H_rho purity bound}). Equation
(\ref{eq: H_rho purity bound}) provides a speed limit to the variation
of the second order Rényi entropy $-\ln\mc P$ \cite{RenyiEntropyQM}
in terms of the Hilbert-Schmidt norm of the Lindblad operators. Its
practical usefulness stems from the fact that the operators $A_{k}$
can be experimentally determined in large variety of quantum systems
\cite{chu97,how06,obr04,rie06,nee08,bru08}. As we shall see this
simple bound is not very tight and scales badly with the number of
levels. Fortunately, this can be remedied by using Liouville space.

\section{Purity speed limit in Liouville space}

Quantum dynamics is traditionally described in Hilbert space. However,
for open quantum systems, it is sometimes convenient to introduce
an alternative space where density operators are represented by vectors,
and time evolution is generated by superoperators that operate on
vectors just from the left (as in Schrödinger equation). This space
is referred to as Liouville space \cite{muk99}. We denote the \char`\"{}density
vector\char`\"{} by $\ket{\rho}\in\mathbb{C}^{1\times N^{2}}$. It
is obtained by reshaping the density matrix $\rho$ into a larger
single vector with index $\alpha\in\{1,2,....N^{2}\}$. The vector
$\ket{\rho}$ is not normalized to unity in general. The norm squared
is equal to the purity, $\mc P=\text{tr}(\rho^{2})=\braket{\rho}{\rho}$
where $\bra{\rho}=\ket{\rho}^{\dagger}$, as usual. From the identity
$id_{t}\rho_{\alpha}=\sum_{\beta}[i\partial(d_{t}\rho_{\alpha})/\partial\rho_{\beta}]\rho_{\beta}=\sum_{\beta}\mc H_{\alpha\beta}\rho_{\beta}$
it follows that in Liouville space the equation of motion (\ref{eq: Lindblad eq.})
takes a Schrödinger-like form, 
\begin{equation}
i\partial_{t}\ket{\rho}=\mc H\ket{\rho},\label{eq: schrodinger eq}
\end{equation}
when using the common matrix to vector index mapping $\alpha=(row-1)N+column$,
the Hamiltonian superoperator $\mc H\in\mathbb{C}^{N^{2}\times N^{2}}$
is given by \cite{vecingMachnesPlenio,Horn} 
\begin{eqnarray}
\mc H & =-i(H\otimes I-I\otimes H^{t})\nonumber \\
 & +i\sum_{k}(A_{k}\otimes A_{k}^{*})-\frac{1}{2}I\otimes(A_{k}^{\dagger}A_{k})^{t}-\frac{1}{2}A_{k}^{\dagger}A_{k}\otimes I\label{eq: Hr form}
\end{eqnarray}
The superoperator $\mc H$ is non-Hermitian for open quantum systems.
The skew Hermitian part $(\mc H-\mc H^{\dagger})/2$ is responsible
for purity changes and stems uniquely from the Lindblad operators
$A_{k}$ of the master equation (\ref{eq: Lindblad eq.}). To derive
a purity speed limit in Liouville space we use (\ref{eq: schrodinger eq}),
and obtain the equality, $\partial_{t}\ln\braket{\rho}{\rho}=-i\braOket{\rho}{\mc H-\mc H^{\dagger}}{\rho}/\braket{\rho}{\rho}$.
Integrating this expression over time and using the triangle inequality,
we get, 
\begin{equation}
\left|\ln\frac{\mc P(t_{f})}{\mc P(t_{i})}\right|\le\int_{t_{i}}^{t_{f}}\frac{\left|\braOket{\rho}{\mc H-\mc H^{\dagger}}{\rho}\right|}{\braket{\rho}{\rho}}dt.\label{eq: r purity evo triang}
\end{equation}
The integrand may be further bounded by the spectral (or operator)
norm 
\begin{equation}
\frac{\left|\braOket{\rho}{\mc H-\mc H^{\dagger}}{\rho}\right|}{\braket{\rho}{\rho}}\le\nrm{\mc H-\mc H^{\dagger}}_{\text{sp}}.\label{eq: spec norm speed bound}
\end{equation}
For skew Hermitian operators like $\mc H-\mc H^{\dagger}$ with eigenvalues
$\lambda_{i}$, the spectral norm is equal to $\text{max}|\lambda_{i}|$
\cite{bha97}. Combining (\ref{eq: r purity evo triang}) and (\ref{eq: spec norm speed bound}),
we eventually obtain a cumulative speed limit for the second order
Réyni entropy of the form (\ref{eq: G cumu}):

\begin{equation}
\left|-\ln\mc P(t_{f})+\ln\mc P(t_{i})\right|\le\int_{t_{i}}^{t_{f}}\nrm{\mc H-\mc H^{\dagger}}_{\text{sp}}dt.\label{eq: Hr purity bound}
\end{equation}
Here as well, instead of a speed limit for $-\ln\mc P$ it is possible
to get a less tighter speed limit for $\mc P$. We point out that
the norm $\nrm{\mc H-\mc H^{\dagger}}_{\text{sp}}$ can be directly
determined from the measurable Lindblad operators $A_{k}$ using (\ref{eq: Hr form}).

\section{Superiority of the Liouville space bound}

Next, we show that the Liouville space speed limit is always tighter
compared to the Hilbert space bound (\ref{eq: H_rho purity bound}).
From (\ref{eq: Hr form}) and the triangle inequality 
\begin{eqnarray*}
\nrm{\mc H-\mc H^{\dagger}}_{sp} & \le\sum_{k}\nrm{A_{k}\otimes A_{k}^{*}}_{sp}+\nrm{A_{k}^{\dagger}\otimes A_{k}^{t}}_{sp}\\
 & +\nrm{I\otimes[(A_{k}^{\dagger}A_{k})^{t}}_{sp}+\nrm{A_{k}^{\dagger}A_{k}\otimes I}_{sp}
\end{eqnarray*}
Using $\nrm{A\otimes B}_{sp}=\nrm A_{sp}\nrm B_{sp}$ and submutiplicativity
of the spectral norm we get 
\begin{equation}
\nrm{\mc H-\mc H^{\dagger}}_{sp}\le4\nrm{A_{k}}_{sp}^{2}\le4\nrm{A_{k}}_{2}^{2}\label{eq: Lio vs Hilb}
\end{equation}
Where the last inequality is a general relation between the spectral
norm and the HS norm \cite{Horn}. Equation (\ref{eq: Lio vs Hilb})
gives two important results. First, it proves our claim that the Liouville
space speed limit is always tighter compared to the Hilbert space
speed limit (\ref{eq: H_rho purity bound}). The second result is
that by replacing $4\nrm{A_{k}}_{2}^{2}$ with $4\nrm{A_{k}}_{sp}^{2}$
in (\ref{eq: spec norm speed bound}), a new tighter bound in Hilbert
space is obtained. On the one hand it uses the original Lindblad operator
$A_{k}$, and on the other hand is always tighter than (\ref{eq: H_rho purity bound}).
In addition it does not suffer from the scaling problems we discuss
next.

\textit{Scaling behavior}. Let us consider $M$ independent particles
subjected to the same dynamics. Since the system is in a product state
at all times, the log-purity scales like $\ln\mc P_{M}\sim M\ln\mc P_{1}$.
The Liouville space bound (\ref{eq: Hr purity bound}) then exhibits
the correct $M$ scaling, $d\mc P_{M}/dt\leq M||\mc H-\mc H^{\dagger}||_{\text{sp}}$,
where $\mc H$ is the single particle super Hamiltonian. In contrast,
the Hilbert space bound (\ref{eq: H_rho purity bound}) is, $d\mc P_{M}/dt\leq M2^{M-1}||A||_{2}^{2}$,
and the left hand side lead to an exponential overestimation of the
speed limit. In addition, for a single particle $N$-level dephasing
channel with eigenvalues $\lambda_{j}(A)=\exp(i\varphi_{j})$, we
find that the purity speed in Liouville space is limited by $\text{max}|\lambda_{i}-\lambda_{j}|^{2}\leq4$,
while the HS norm that appears in Hilbert space speed limit is equal
to $4N$. Hence the Hilbert space bound (\ref{eq: H_rho purity bound})
dramatically overestimating the purity value for large $N$. Note
that the scaling problems are resolved if the HS norm is replaced
by the spectral norm (equation (\ref{eq: Lio vs Hilb}) shows why
this is legitimate to do).

\section{Purity deviation}

In what follows we introduce the purity deviation. It has two main
advantages over the regular purity. The first is that it enables to
get a speed limit which is even tighter than (\ref{eq: Hr purity bound}).
The second advantage is that it provides information that is sometimes
more useful than the purity. Essentially it quantifies the distance
between two different solution of the system as a function of time.
The is very useful in studying relaxation to steady state dynamics.
Let $\ket{\rho_{s}}$ be a specific solution of the quantum evolution
$i\partial_{t}\ket{\rho_{s}}=\mc H\ket{\rho_{s}}$. We define the
deviation vector as $\ket{\rho_{D}}=\ket{\rho}-\ket{\rho_{s}}$, and
the corresponding purity deviation as $\mc P_{D}=\braket{\rho_{D}}{\rho_{D}}$.
The purity deviation has a simple geometrical meaning as the square
of the Euclidean distance, $\text{tr}[(\rho-\rho_{s})^{2}]$, between
the states $\rho$ and $\rho_{s}$ (the regular purity is the distance
to the origin $\rho_{s}=0$). By taking the time derivative of $\mc P_{D}$
and repeating the previous derivation, we find 
\begin{equation}
\left|\ln\frac{\mc P_{D}(t_{f})}{\mc P_{D}(t_{i})}\right|\le\int_{t_{i}}^{t_{f}}\nrm{\mc H-\mc H^{\dagger}}_{\text{sp}}dt,\label{eq: P dev bound}
\end{equation}
where the purity $\mc P$ has now been replaced by the purity deviation
$\mc P_{D}$. While (\ref{eq: P dev bound}) is valid for all vectors
$\ket{\rho_{s}}$ that solves \ref{eq: schrodinger eq}, it becomes
particularly useful when $\ket{\rho_{s}}$ is given by the steady
state, $i\partial_{t}\ket{\rho_{s}}=0$. The benefit of the replacement
$\mc P\rightarrow\mc P_{D}$ is that only the part of the purity that
changes in time is taken into account. The purity deviation bound
(\ref{eq: P dev bound}) has the remarkable property that it may be
tight at all times for purely dephasing qubit channels and for the
erasure of classical correlations (see below).

\begin{figure}
\includegraphics[width=8.4cm]{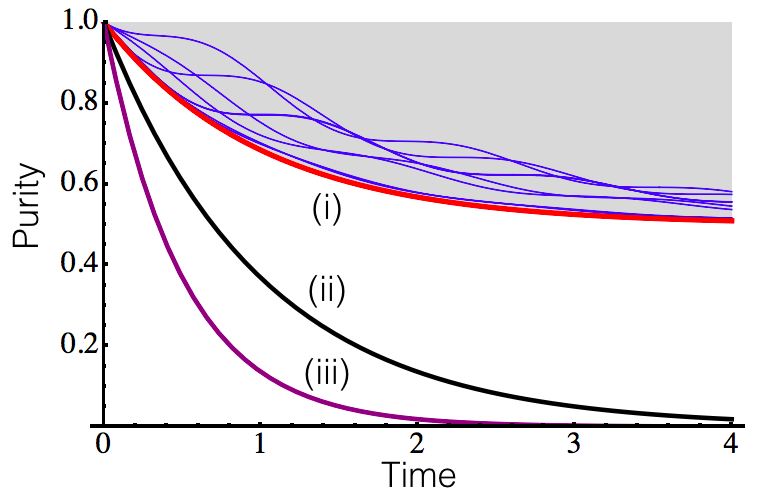}

\protect\protect\caption{Purity bounds for a qubit dephasing channel with $H_{\rho}=\sigma_{x}$
and $A=\sigma_{z}$ in a pure state at $t_{i}=0$. The bound (\ref{12})
(curve i) based on the purity deviation bound (\ref{eq: P dev bound})
is tighter than the purity bound in Liouville space (\ref{eq: Hr purity bound})
(curve ii), and the corresponding bound in Hilbert space (\ref{eq: H_rho purity bound})
(curve iii). The bound (12) clearly delimits the region of allowed
purities (blue lines obtained for various random initial conditions)
and is thus tight.}
\end{figure}

\section{Applications}

(a) Dephasing channel. To illustrate the strength of the purity deviation
bound even for the simplest scenarios, we first apply it to a pure
dephasing of a two-level qubit. The system is described by $H=\sigma_{z}$
and one nonzero Lindblad operator that satisfies $[A,A^{\dagger}]=0$
and $[A,H]=0$ \cite{nie00,LidarAlickiPurityDecrease}. Without loss
of generality, we assume that the operator $A$ is traceless \cite{bre07}.
In this situation, the Hilbert space bound (\ref{eq: H_rho purity bound})
takes a minimal value that is exactly two times larger than the tighter
Liouville space bound (\ref{eq: Hr purity bound}). For instance,
for $A=\sigma_{z}$ and an initial density matrix of the form $\rho(t_{i})=\{\{a,b\},\{b^{*},1-a\}\}$,
we find $|\ln\mc P(t_{f})/\mc P(t_{i})|\le2(t_{f}-t_{i})$ in Hilbert
space and $|\ln[\mc P(t_{f})/\mc P(t_{i})]|\le t_{f}-t_{i}$ in Liouville
space. Remarkably, the purity deviation bound (\ref{eq: P dev bound})
is tight at all times in this case: we choose $\rho_{s}$ to be the
steady state given by the fully mixed state $\rho_{s}=\{\{a,0\},\{0,1-a\}\}$
(there may be several steady states, and one may choose one of them,depending
on the initial state), and obtain the equality $|\ln[\mc P_{D}(t_{f})/\mc P_{D}(t_{i})]|=|\ln[2b^{2}e^{-t_{f}}/(2b^{2}e^{-t_{i}})]|=t_{f}-t_{i}$
which is exactly equal to the right hand side of (\ref{eq: P dev bound}).
As will be shown later, the purity deviation can lead to a tight speed
limit even in multi-particle scenarios with quantum correlations.

In general, the purity deviation leads to a better bound on the purity
change for a general dephasing channel in a $N$-level system. Choosing
$\rho_{s}$ to be the fully mixed state (which always corresponds
to a steady state in dephasing dynamics) yields $\text{tr}[(\rho-\rho_{s})^{2}]=\text{tr}[\rho^{2}]-1/N$.
Using this in Eq. (\ref{eq: P dev bound}), we obtain, 
\begin{equation}
\mc P(t_{f})\ge\frac{1}{N}+\left(\mc P(t_{i})-\frac{1}{N}\right)\exp\left[-\int_{t_{i}}^{t_{f}}\nrm{\mc H-\mc H^{\dagger}}_{\text{sp}}dt\right].\label{12}
\end{equation}
This equation, valid for any dephasing channel, is always better compared
to Eq. (\ref{eq: Hr purity bound}). Figure 1 shows the purity (blue
lines) for pure random initial states in a simple one qubit dephasing
channel. The purity bound (\ref{12}) derived from the purity deviation
bound (\ref{eq: P dev bound}) is significantly better than the other
two bounds, and tightly delimits the regime of allowed purity values.

(b) Quantum correlation erasure in $N$-particle system. Quantum correlations
are subject of intense study in quantum information theory as well
as in the collective dynamics of multi-particle systems. In quantum
information quantum correlations are a resource that can be exploited
for computational speedup. It is of interest to evaluate how fast
this information dissipates away (either intentionally or unintentionally).
Consider a chain of $M$ qubits that are not in a product state. This
state may contain both quantum and classical correlations between
the different particles. In this example we study the speed limit
for erasing quantum correlation while leaving the classical correlations
(in the basis of interest) intact. As shown next, this happens naturally
in the presence of dephasing. Let, $\{\ket 0,\ket 1\}^{\otimes M}$
be the some basis of interest (e.g. the energy basis) in which the
erasure takes place. Given some initial density matrix of the whole
$N$-particle system in this basis $\rho_{i}$, the corresponding
state with no quantum correlation is $\rho_{f}=Diagonal(\rho_{i})$.
To quantify the speed at which we approach the quantum correlation
free state $\rho_{f}$ we use the standard Hilbert Schmidt norm 
\begin{equation}
R(t)=tr[(\rho(t)-\rho_{f})^{2}].
\end{equation}
This quantity has the structure of purity deviation since both $\rho(t)$
and $\rho_{f}$ are valid solution of the equation of motion. In order
to remove quantum correlation a dephasing operation is needed. This
dephasing can be achieved by two different means. Either by local
dephasing on each particle or by global dephasing operators. We start
from local dephasing by applying $M$ Lindblad terms of the form $A_{1}=\sqrt{\gamma_{local}}\sigma_{z}\otimes I\otimes I..,\:A_{2}=\sqrt{\gamma_{local}}I\otimes\sigma_{z}\otimes I\otimes I..$
and so on. From (\ref{eq: Hr form}) one can verify that $\nrm{\mc H-\mc H^{\dagger}}=2M\left|\gamma_{local}\right|$
and that the maximal rate is achieved for a GHZ (\cite{GHZ1}) pure
states: $\frac{1}{\sqrt{2}}(\ket{a,b,c,...}+\ket{1-a,1-b,1-c,...})$
where $a,b,c\in\{0,1\}$. Using (\ref{eq: P dev bound}), the speed
limit for local erasure of quantum correlation is 
\begin{equation}
R(t)\ge R(0)e^{-\gamma_{local}Mt}\label{eq: R local deph}
\end{equation}
where \textit{the speed limit is tight for GHZ states.} Next we consider
a global dephasing operator of the form $\{A_{m,n}^{(k)}\}_{k=1}^{M}=\sqrt{\gamma_{global}}\delta_{m,k}\delta_{n,k}$.
Using (\ref{eq: Hr form}) one can verify that $\mc H-\mc H^{\dagger}$
has $M$ singular values that are equal to zero and the rest are equal
to $\left|\gamma_{global}\right|$ and therefore 
\begin{equation}
R(t)=R(0)e^{-\gamma_{global}t}
\end{equation}
Remarkably, the bound yields an \textit{exact tight result for any
initial state} (since all the modes (coherences) decay at the same
rate). These two different erasure processes not only show the difference
between local and global erasure but also demonstrate that \textit{the
speed limits we derived based on Liouville apace and the purity deviation,
can be tight for arbitrarily large and entangled system}.

(c) Interacting particles. To demonstrate that our bound is also applicable
for systems with strongly interacting particles, we consider the case
of local dephasing for each particle in a chain of $M$ interacting
particles. This is exactly the scenario in recent studies of heat
transport in ion chains \cite{dephase_ion_chain,ion_chain_exp}. Since
the fully mixed state is a solution for any type of interaction we
obtain that (\ref{eq: R local deph}) holds where this time $R(t)=tr(\rho-\rho_{I})$
where $\rho_{I}$ is the density matrix of the fully mixed M-particle
state. 
\begin{figure}
\includegraphics[width=8.6cm]{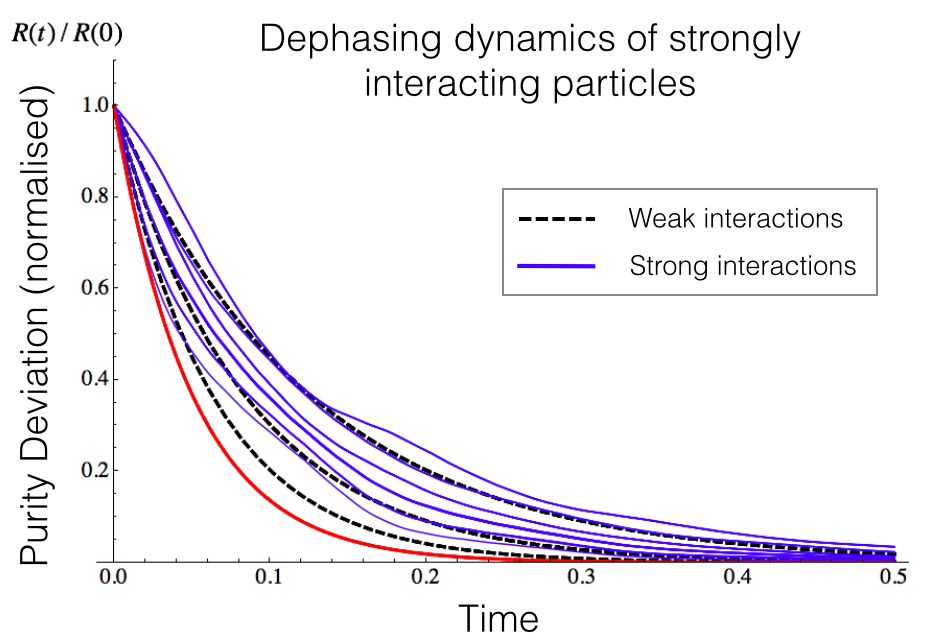}

\caption{Normalized purity deviation $R(t)/R(0)$ (see example (c)) of five
qubit particles with nearest neighbor interaction and local dephasing
channels. The black-dashed (solid-blue) lines are numerical result
for random initial states and weak (strong) interactions. The red
line shows the purity deviation speed limit we derived. It is observed
that the results hardly depend on the number of particles. }
\end{figure}

In Fig. 2 we plot $R(t)/R(0)$ for five interacting particles with
time-dependent nearest neighbor interaction of the form $V_{i,i+1}=V_{0}\cos(t)\sigma_{x}^{i}\otimes\sigma_{x}^{i+1}.$
The Hamiltonian particle $'i'$ is $H_{i}=\sigma_{z}$ and the local
dephasing operator of each particle is $A_{i}=\sigma_{z}$. The red
line is our bound on R(t)/R(0), the black-dashed lines are numerical
results for random initial condition with $V_{0}=0.1$, and the solid-blue
lines are numerical results with strong interaction $V_{0}=10$.

(d) Erasing classical and quantum correlations by resetting a subsystem.
The creation of quantum correlations between different systems \cite{nie00}
is a key ingredient in quantum information. Recently, however, there
is a growing interest in the converse problem of correlation erasure.
That is, the removal of correlations between two (or more) systems,
while leaving the local information (the reduced density operators)
intact \cite{Terno,mor99,mor99a,ari07,ari08,dup14}. The problem of
quantum state decorrelation (also called quantum decoupling or disentanglement)
plays an important role in quantum state merging \cite{hor05}, the
computing of channel capacities \cite{hay08}, quantum cryptography
\cite{CryptDecor}, as well as in the study of thermalization \cite{RelTherm}.
Experimental realizations of decorrelation have been reported in \cite{bou99,tek01}.
We shall use our approach to determine the maximal rate of correlation
erasure. This example serves two purposes. The first is to show an
application of our theory to quantum information. The second purpose
is to show that even in cases where the purity deviation cannot be
explicitly related to the standard purity, it can still carry valuable
information that does not appear in the standard purity.

Consider two systems $A$ and $B$ that are initially correlated (correlations
may be classical and/or quantum). The joint density matrix is denoted
by $\rho_{AB}$ and the respective reduced density matrices are $\rho_{A(B)}=\text{tr}_{B(A)}\rho_{AB}$.
We wish to decorrelate $A$ from $B$ by coupling it weakly to a Markovian
reservoir (we assume that there is no interaction between $A$ and
$B$ during the decorrelation process). The corresponding quantum
decorrelator generates the transformation, 
\begin{equation}
\rho_{AB}\to\rho_{A}\otimes\rho_{0},\label{eq: gen decor trans}
\end{equation}
where $\rho_{0}$ is some predetermined state that is independent
of $\rho_{AB}$. Ideally, one would wish the final state to be $\rho_{A}\otimes\rho_{B}$
for any $\rho_{AB}$. However, this operation was shown to be nonlinear
in general \cite{Terno}. In cases where $\rho_{B}$ is known, one
may choose $\rho_{0}=\rho_{B}$. However, this would lead to a decorrelator
that is tuned to a specific $\rho_{B}$.

We use the standard $L2$ norm to define the distance between the
initial correlated state $\rho_{AB}$ and the final decorrelated state
$\rho_{A}\otimes\rho_{0}$, 
\begin{equation}
R_{\text{dec}}=\text{tr}[(\rho_{AB}-\rho_{A}\otimes\rho_{0})^{2}].\label{eq: R distance def}
\end{equation}
If $\rho_{0}=\rho_{B}$ this measure is very similar to the geometric
discord introduced in \cite{GeomDiscord}; However, $R_{\text{dec}}$
describes the distance to a state where system A has neither quantum
nor classical correlation to system B. Demanding that $\rho_{A}\otimes\rho_{0}$
is a solution of the equation of motion Equation (\ref{eq: R distance def})
takes the form of a purity deviation. Next we show that the one-partite
Lindblad operator $L_{\text{dec}}=1_{A}\otimes L_{B}$, where $L_{B}$
has a single stationary state $\rho_{0}$, achieves the decorrelating
transformation given in \eqref{eq: gen decor trans}. After showing
that, we shall apply our purity deviation speed limit to understand
how fast $R_{\text{dec}}$ decreases and $A$ becomes decorrelated
from $B$.

Consider the Lindblad equation of motion that describes an interaction
with a bath in the Markovian limit

\begin{eqnarray}
\partial_{t}\rho_{AB} & = & L[\rho_{AB}]
\end{eqnarray}
where the Lindblad form of $L$ is given in (\ref{eq: Lindblad eq.}).
To leave system $A$ intact we choose the form: 
\begin{eqnarray}
L_{\text{dec}} & = & 1_{A}\otimes L_{B},\\
L_{B}[\rho_{0}] & = & 0,
\end{eqnarray}
where $\rho_{0}$ is the only zero state of $L_{B}$: $L_{B}[\rho]=0\iff\rho=\rho_{0}$.
Since the Lindblad operator has a tensor product form we can write
the evolution as: 
\begin{equation}
\rho_{AB}(t)=e^{L_{\text{dec}}t}[\rho_{AB}(0)]=I_{N\times N}\otimes e^{tL_{B}}[\rho_{AB}(0)].
\end{equation}
The only steady state of $K_{B}=e^{tL_{B}}$ is $\rho_{0}$ and therefore
we can write: 
\begin{equation}
\lim_{t\to\infty}K_{B}[\sigma]=\text{tr}(\sigma)\rho_{0}\label{eq: steady state}
\end{equation}
where we considered the slightly more general case where the initial
trace is different from one (the Lindblad map conserves the trace).
To show how $L_{\text{dec}}$ operates on a general density matrix
we decompose the initial density matrix in the following way:

\begin{align}
\rho_{AB} & =\frac{1}{N^{2}}[I_{N^{2}\times N^{2}}+\sum_{i=1}^{N_{Z}}r_{A,i}Z_{i}\otimes I_{N\times N}+\nonumber \\
 & \sum_{i=1}^{N_{Z}}r_{B,i}I_{N\times N}\otimes Z_{i}+\sum_{i,j=1}^{N_{Z}}t_{ij}Z_{i}\otimes Z_{j}],\label{eq: rho init}
\end{align}
where $N_{z}=N^{2}-N+1$, and $Z_{i}$ are \textit{traceless} orthonormal
basis operators for $N\times N$ Hermitian traceless matrices. $r_{A}$
and $r_{B}$ determine the reduced density matrices: 
\begin{equation}
\text{tr}_{B}\rho_{AB}=\frac{1}{N}[I_{N\times N}+\sum r_{A,i}Z_{i}]=\rho_{A}.
\end{equation}
Next we use \eqref{eq: steady state}, and apply $\lim_{t\to\infty}K$
to the initial state \eqref{eq: rho init} and get

\begin{equation}
\rho_{AB}(t\to\infty)=\frac{1}{N}[I_{N\times N}\otimes\rho_{0}+\sum r_{A,i}Z_{i}\otimes\rho_{0}]=\rho_{A}\otimes\rho_{0},
\end{equation}
which show that $L_{\text{dec}}$ realizes the decorrelating transformation
$\rho_{AB}\to\rho_{A}\otimes\rho_{0}.$ An immediate conclusion from
this result follows: even when removing the correlation is not the
objective, it occurs naturally when the dynamics is generated by a
one-party (local) single-steady-state Markovian map.

Now that we have established $\rho_{AB}\to\rho_{A}\otimes\rho_{0}$
we readily apply (\ref{eq: P dev bound}), and get that the norm integral
needed to change the decorrelation purity from $R_{\text{dec,i}}$
to $R_{\text{dec,f}}$ in time $T$ is, 
\begin{equation}
\left|\ln\frac{R_{\text{dec,f}}}{R_{\text{dec,i}}}\right|\le\int_{0}^{T}\nrm{\mc H_{B}-\mc H_{B}^{\dagger}}dt,\label{15}
\end{equation}
where $\mc H_{B}$ is the Liouville space representation of $L_{B}$.

\begin{figure}
\includegraphics[width=8.7cm]{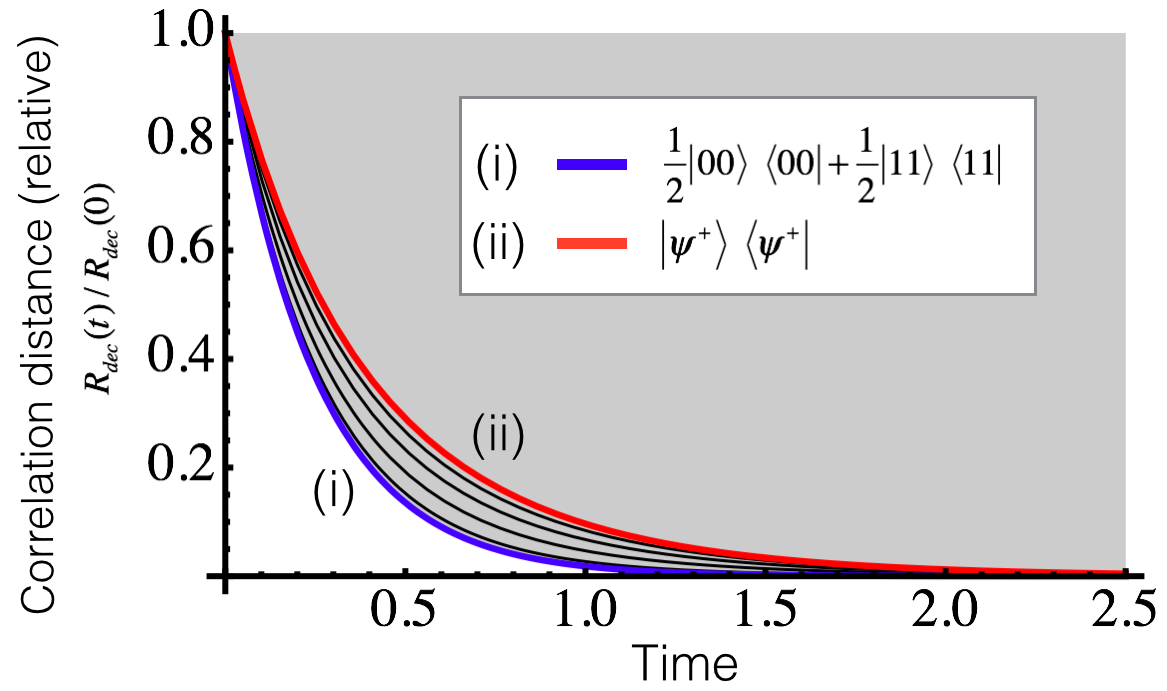}

\protect\caption{(Black) Correlation distance \eqref{eq: R distance def} as a function
of time for the initial correlated state $\rho_{AB}=(\lambda/2)I_{2\times2}+(1-\lambda)\protect\ketbra{\Psi^{+}}{\Psi^{+}}$,
for different values of $\lambda$. The derived decorrelation speed
\eqref{15} (gray area) delimits the regime of accessible correlation
values as a function of time. The erasure of classical correlations
(curve i) ($\lambda=1$) occurs here faster than the erasure of quantum
correlations (curve ii) ($\lambda=0$). The decorrelation speed is
tight as it is attained in this case by the classical correlated state.}
\end{figure}

We illustrate the usefulness of \eqref{15} by considering the initial
state $\rho_{AB}=(\lambda/2)I_{2\times2}+(1-\lambda)\ketbra{\Psi^{+}}{\Psi^{+}}$
where $I_{2\times2}$ is the unit operator, $\ket{\Psi^{+}}$ the
usual Bell state and $0\le\lambda\le1$. The reduced density operators
are simply $\rho_{A}=\rho_{B}=\frac{1}{2}I_{2\times2}$. The quantum
discord of the joint state is monotonically increasing from zero to
one as $\alpha$ changes from zero to one \cite{LuoDiscord}. We take
the quantum decorrelator $L_{\text{dec}}$ as the sum of the Lindblad
operators $I_{2\times2}\otimes\sigma_{\pm}$ where $\sigma_{\pm}$
are the raising and lowering operators in spin space. This is equivalent
to infinite temperature bath (or in practice, a bath whose temperature
is much larger than the gap of the system). Figure 2 shows the relative
decorrelation distance, $R_{\text{dec}}(t)/R_{\text{dec}}(0)$, as
a function of time for different values of the parameter $\lambda$.
Interestingly, in this case, classical correlations (blue) are erased
faster than quantum correlations (red). In fact the classical case
constitutes an example where the decorrelation speed bound (\ref{15})
is tight.

Finally we point that the decorrelation process described here is
quite different from simply replacing system B with a new system that
is in a state $\rho_{0}$ (that is not correlated to system A). In
this case nothing is accomplished since system A is still fully correlated
to the old system B. However, the Lindblad dynamics we use to decorrelate
describes weak coupling to the environment. Thus, in the dynamics
above the correlation is spread over the vast number of degrees of
freedom in the bath. Consequently it is effectively lost and cannot
be retrieved. The slightest noise will make it impossible to get the
correlations back by time reversal.

\textit{Conclusions.} The Liouville space approach was used to derive
a global \textit{state-independent} quantum speed limit for purity
changes that is much tighter than the speed limit obtained by using
the standard Hilbert space approach. Furthermore, by introducing the
purity deviation with respect to the steady-state of the system, an
even more accurate speed limit was obtained. Remarkably, the purity
deviation bound can be attained, and therefore constitutes a tight
speed limit. We point out that the same techniques can be applied
to the von Neumann entropy and other Rényi entropies. To demonstrate
the utility of our results to quantum information theory and its applications,
we derived speed limits for multi-particle dephasing processes (including
inter-particle interactions) and for classical and quantum correlation
erasure. Due to its versatility and its compact-easy to use form,
we expect the purity speed limit to become a useful tool in the investigation
of the dynamics of open quantum systems, from coherent control and
unitary gates implementation to quantum thermodynamics.

\begin{acknowledgments}
The authors thank Eric Lutz for useful tips and comments. This work
was supported by the Israeli science foundation
\end{acknowledgments}

\bibliographystyle{apsrev4-1}
\bibliography{testbib}

\end{document}